%% file: main.tex
\documentclass[10pt,conference]{IEEEtran} 
\IEEEoverridecommandlockouts
\usepackage{cite}
\usepackage{amsmath,amssymb,amsfonts}
\usepackage{algorithmic}
\usepackage{graphicx}
\usepackage{textcomp}
\usepackage{xcolor}
\usepackage{xspace}
\usepackage{booktabs}
\usepackage[a4paper, total={184mm,239mm}]{geometry}
\usepackage[labelformat=simple]{subcaption}
\usepackage[font=small,labelfont=bf]{caption}

\def\BibTeX{{\rm B\kern-.05em{\sc i\kern-.025em b}\kern-.08em
    T\kern-.1667em\lower.7ex\hbox{E}\kern-.125emX}}

\newcommand{\wifi}{Wi-Fi\xspace}
\newcommand{\noble}{NObLe\xspace}
\newcommand{\UJI}{UJIIndoorLoc\xspace}

\begin{document}

\title{Neighbor Oblivious Learning (\noble) for Device Localization and Tracking\\
}

\author{\IEEEauthorblockN{ Zichang Liu}
\IEEEauthorblockA{\textit{Department of Computer Science} \\
\textit{Rice University}\\
Houston, USA \\
zichangliu@rice.edu}
\and
\IEEEauthorblockN{Li Chou}
\IEEEauthorblockA{\textit{Department of Computer Science} \\
\textit{Rice University}\\
Houston, USA \\
lchou@rice.edu}
\and
\IEEEauthorblockN{ Anshumali Shrivastava}
\IEEEauthorblockA{\textit{Department of Computer Science} \\
\textit{Rice University}\\
Houston, USA \\
anshumali@rice.edu}
}

\maketitle

\input{abstract.tex}

\input{introduction.tex}

\input{background.tex}

\input{model.tex}

\input{Wifiexperiment.tex}

\input{IMUExperiment.tex}

\input{conclusion.tex}
\input{localization-bib}

\end{document}

%% file: abstract.tex
\begin{abstract}
On-device localization and tracking are increasingly crucial for various applications. Along with a rapidly growing amount of location data, machine learning (ML) techniques are becoming widely adopted. A key reason is that ML inference is significantly more energy-efficient than GPS query at comparable accuracy, and GPS signals can become extremely unreliable for specific scenarios. To this end, several techniques such as deep neural networks have been proposed. However, during training, almost none of them incorporate the known structural information such as floor plan, which can be especially useful in indoor or other structured environments. In this paper, we argue that the state-of-the-art-systems are significantly worse in terms of accuracy because they are incapable of utilizing this essential structural information. The problem is incredibly hard because the structural properties are not explicitly available, making most structural learning approaches inapplicable. Given that both input and output space potentially contain rich structures, we study our method through the intuitions from manifold-projection. Whereas existing manifold based learning methods actively utilized neighborhood information, such as Euclidean distances, our approach performs Neighbor Oblivious Learning (NObLe). We demonstrate our approach's effectiveness on two orthogonal applications, including \wifi-based fingerprint localization and inertial measurement unit(IMU) based device tracking, and show that it gives significant improvement over state-of-art prediction accuracy. 

\end{abstract}

%% file: introduction.tex
\section{Introduction}
The global market size for location-based services is expected to grow to USD 26.7 billion by 2025 from USD 13.8 billion in 2020 \cite{globalfact}. The key to the projected growth is an essential need for accurate location information. For example, location intelligence is critical during public health emergencies, such as the current COVID-19 pandemic, where governments need to identify infection sources and spread patterns. Traditional localization systems rely on global positioning system (GPS) signals as their source of information. However, GPS can be inaccurate in indoor environments and among skyscrapers because of signal degradation. Moreover, GPS is notorious for battery drainage because of slow and demanding communication requirements \cite{TO-Improving}. Therefore, GPS alternatives with higher precision and lower energy consumption are urged by industry. Existing network infrastructure such as \wifi(IEEE 802.11) is utilized for localization \cite{RADAR}\cite{AL-wifi} to avoid expensive infrastructure deployment. Besides, low-cost inertial measurement sensors (IMU) based on accelerators and gyroscopes, which are widely embedded in modern mobile devices, have also emerged as popular solution \cite{JY-IMU}\cite{QY-IMU} for both indoor and outdoor device tracking task. An informative and robust estimation of position based on these noisy inputs would further minimize localization error.

Machine learning (ML) techniques are a logical choice for these estimation tasks, and popular algorithms such as $k$-nearest neighbors and random forest have been proposed \cite{HORUS}\cite{Chen_date}.  Since deep neural networks (DNN) have performed surprisingly well in computer vision, natural language processing, and information retrieval, many attempts have been made to utilize DNNs for localization \cite{DeepFi}\cite{WiDeep}\cite{CNNLoc}. These approaches either formulate localization optimization as minimizing distance errors or use deep learning as denoising techniques for more robust signal features.

\vspace{-8mm}
\begin{figure}[htbp]
\centering
 	\begin{subfigure}[t]{0.2\textwidth}
 	    \centering
 	    \includegraphics[trim=0 40 0 0,clip,scale=.2]{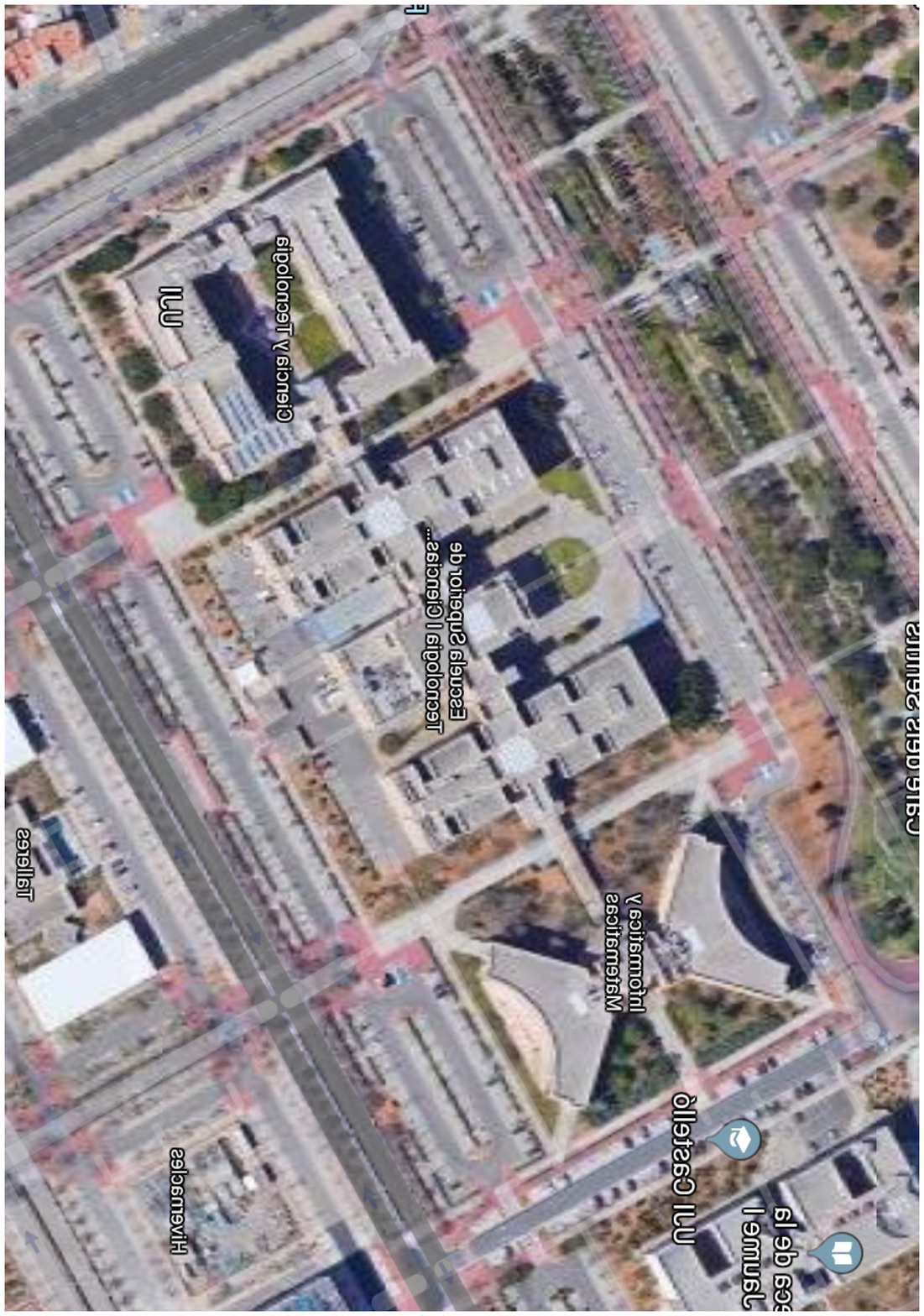}
     \end{subfigure}
 	\begin{subfigure}[t]{0.2\textwidth}
 	    \centering
 	    \includegraphics[trim=60 60 60 60,clip,width=3.5cm,height=5cm]{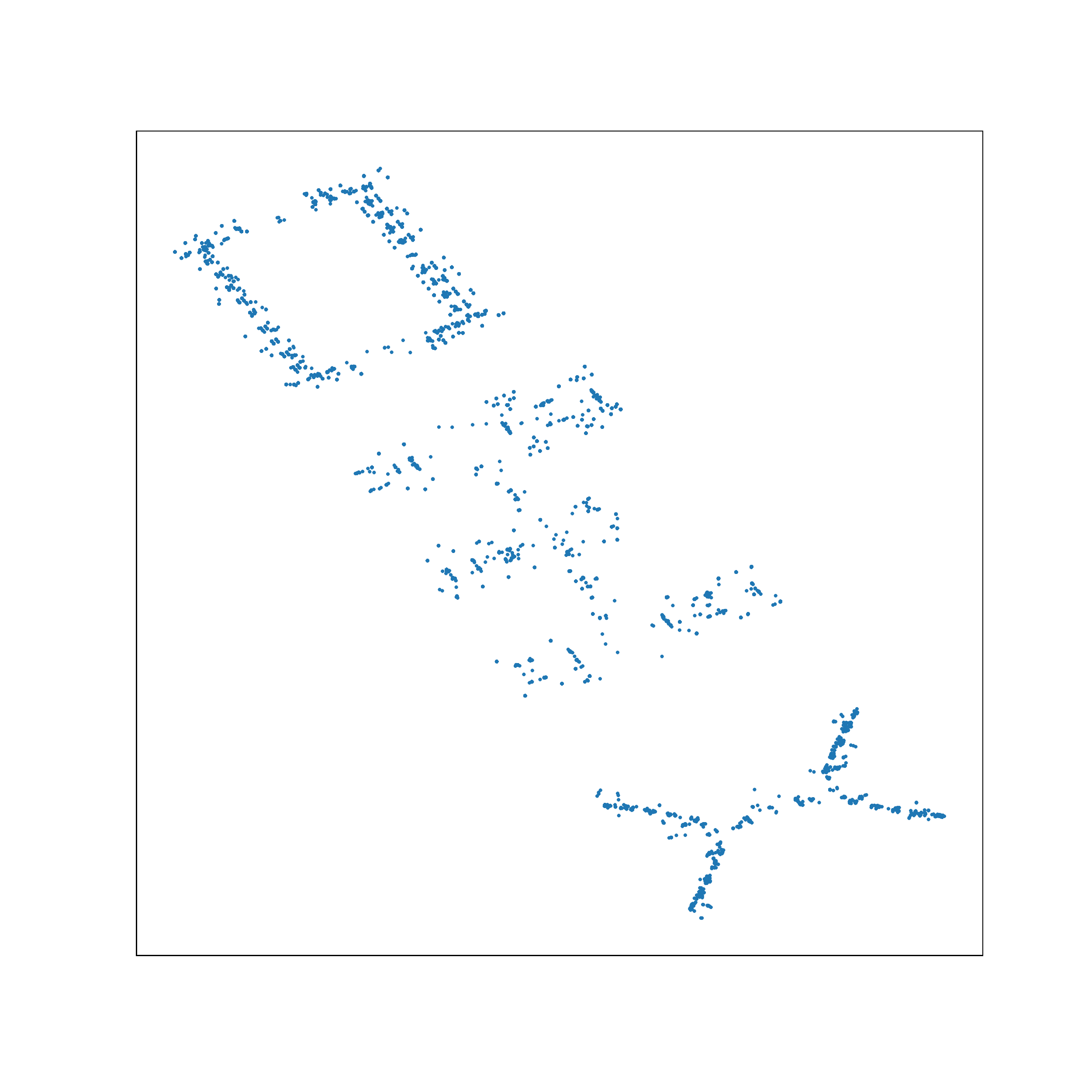}
     \end{subfigure}
     \caption{Both figures corresponds to the three building in \UJI dataset. Left figure is the screenshot of aerial satellite view of the buildings (source: Google Map). Right figure shows the ground truth coordinates from offline collected data.}
     \label{fig:views-intro}
\end{figure}
\vspace{-4mm}
All the methods mentioned above fail to utilize common knowledge: space is usually highly structured. Modern city planning defined all roads and blocks based on specific rules, and human motions usually follow these structures. Indoor space is structured by its design floor plan, and a significant portion of indoor space is not accessible. As an illustration, consider Fig. \ref{fig:views-intro} based on the largest publicly available indoor \wifi localization dataset \UJI\cite{UJILOC}, which covers three buildings with four floors, a space of 397 meters by 273 meters. Space structure is clear from the satellite view, and offline signal collecting locations exhibit the same structure. Fig. \ref{fig:wifi-reg-pred} shows the outputs of a DNN that is trained using mean squared error to map \wifi signals to location coordinates. This regression model can predict locations outside of buildings, which is not surprising as it is entirely ignorant of the output space structure. It was observed in \cite{Chen_date} \cite{7917859} that projecting the predicted outputs to the closest positions on the map would increase localization precision. Our experiment shows that forcing the prediction to lie on the map only gives marginal improvements. In contrast, Fig. \ref{fig:wifi-cls-pred} shows the output of our \noble model, and it is clear that its outputs have a sharper resemblance to the building structures.

We view localization space as a manifold and our problem can be regarded as the task of learning a regression model in which the input and output lie on an unknown manifold. The high-level idea behind manifold learning is to learn an embedding, of either an input or output space, where the distance between learned embedding is an approximation to the manifold structure. In scenarios when we do not have explicit (or it is prohibitively expensive to compute) manifold distances, different learning approaches use nearest neighbors search over the data samples, based on the Euclidean distance, as a proxy for measuring the closeness among points on the actual manifold. While this is justified because the definition of manifold states that any manifold locally is a Euclidean space, however, the Euclidean distances between data pairs may not be a good criterion for finding neighbors on manifold structures for localization services as input signals are extremely noisy. 

{\bf Our Contributions:} Our proposal is inspired by approaches in manifold learning. We argue that Euclidean distance is not reliable for local manifold structure approximation in localization, and propose to ignore small changes in the Euclidean distance and focus on the relative closeness of reconstructed embedding. We propose {\bf N}{\em eighbor} {\bf Ob}{\em livious} {\bf Le}{\em arning} ({\bf \noble}), a DNN approach that achieves structure-aware localization.
Further, we demonstrate the applicability of our techniques on two independent applications: (1) \wifi signal strength based indoor localization and (2) IMU-based device positioning in an outdoor environment. Our evaluations on both applications show that \noble gives significant accuracy improvements. To illustrate that our system can be deployed on energy and computation constraints mobile devices, we thoroughly ran energy tests on two systems. We demonstrate that our model has significantly smaller energy consumption (specifically, 27 times less energy on IMU tracking) than GPS measurements.

%% file: background.tex
\vspace{-1mm}
\section{Background and Related Work}
\vspace{-1mm}
\textbf{Manifold Learning:} Manifold learning is a class of non-linear dimensionality reduction methods. The objective is to find a low-dimensional representation describing some given high-dimensional data observed from an input or feature space $\mathcal{X}$. It is generally assumed that $\forall x \in \mathcal{X}$, $x$ is sampled from some smooth $p$-dimensional submanifold $\mathcal{M} \subset \mathbb{R}^d$. The manifold learning task is to then find a mapping $\psi : x \rightarrow z \in \mathbb{R}^s$ such that $p \leq s \ll d$, while, loosely stated, preserving (structural) properties (e.g., interpoint distances) of the original feature space. Two popular manifold learning methods are locally linear embedding (LLE) \cite{LLE} and isometric mapping (Isomap) \cite{Isomap}. These algorithms follow a template comprised of three steps: (1) construct a neighborhood graph, which involves (expensive) nearest neighbor search; (2) construct a (positive semi-definite) kernel, which is specified as shortest path distances for Isomap, and weights (or coefficients) from solving a system of linear equations for LLE; and (3) perform partial Eigenvalue decomposition.

\textbf{\wifi Localization:} It is cost-effective to leverage existing wireless infrastructure to develop localization techniques. Combining \wifi with radio map is also known as fingerprinting, which consists of two phases. Offline phase: signal features are sampled at selected locations and processed to build the radio map, a database of locations, and their corresponding signal values. One type of signal feature used is received signal strength indicator (RSSI) values from multiple wireless access points (WAP). Online phase: observed RSSI values are matched with points on the radio map to determine the current location, which relies on searching for the most similar locations based on the stored RSSI values in the radio map. 
Many of these techniques do not use structural information.
   

\textbf{Localization on IMU:} Cheap inertial-based sensors on mobile computing devices have emerged as a potential solution for infrastructure-free indoor localization and navigation. However, there are two main challenges. First, IMUs are extremely noisy, making it impossible to use only through physical principles and numerical integration. Second, it keeps updating previous positions, which makes it subject to error accumulation. Various techniques have been proposed to mitigate error accumulation by ruling out illegal movements. A line of work utilizes a floor map to hand-design heuristic rules to correct localization error. For example, \cite{Chen_date} achieved a mean error of 4.3m on a testbed of 163m by 62m. With a map, it uses high-accuracy turn detection to correct positioning error based on the assumption that turns can only be made on specific points on the map.


\textbf{ML in Localization:} Several ML algorithms, such as support vector machines and neural networks, have been applied to localization. Typically, signal strength readings are used as inputs, and outputs are either two or three dimension vectors, corresponding to 2-D or 3-D location estimates~\cite{4343996}. This approach formulates localization as a regression problem that predicts two continuously coordinate variable values given signal strength vector. ML is also used for denoising in order to extract core features for wireless signals. WiDeep \cite{WiDeep} utilize one auto-encoder (AE) for every WAP, making it hard to scale. DeepFi \cite{DeepFi} also utilizes DNNs, but also ignore structure information. CNNLoc \cite{CNNLoc} utilizes a complex architecture including stacked AEs and convolutional neural networks to achieve a mean error of 11.78m on \UJI. ML was also applied to IMU-based localization. \cite{Chen_date} used nearest neighbors and random forest regression to predict the travel distance based on IMU readings.

%% file: model.tex
\vspace{-2mm}
\section{Proposed System Design}
\vspace{-2mm}
\subsection{Intuition}

The world we live in contains many structural themes and elements.  Factoring in structure information usually lead to performance improvement. For example, in computer vision, many state-of-art approaches exploit structure within images. Given the {\em structural} nature of localization space, we approach the problem with the intuition and consideration that the input and output space lies in a manifold space.

Manifold-based learning algorithms, usually unsupervised, utilize local Euclidean distances to approximate neighborhood structure. However, the input features for localization problems are noisy signals. When a person is walking, the accelerometer and gyroscope sensors are likely to pick up a lot of noise due to spurious movements. Moreover, different individuals have different walking styles. Similarly, \wifi signals can be noisy because of moving crowds or room set-ups. Thus, small changes in such noisy input signals are not reliable information about the manifold structure and direct adopting traditional manifold learning approaches is not appropriate. To combat this noise, we ignore small Euclidean differences and propose Neighbor Oblivious Learning (\noble). We propose to quantize the continuous output space into a set of grid-like neighborhood areas, and all data points within the same grid are considered belonging to the same class. It is widely accepted that the penultimate layer of deep neural network classifier model can be regarded as learned embedding~\cite{MM-embedding}~\cite{representation_learning}. We use  DNN and optimize it with cross entropy loss to maximize the embedding distance between different classes, while oblivious to embedding distance within the same class.


\vspace{-2mm}
\subsection{Space Quantization and Multi-label Classification}

Consider a space $S$ for localization. We collect data samples of the form $(\vec{s}, (x, y) )$, where $\vec{s}$ is a vector representing signal features, and $(x, y)$ denotes longitude and latitude coordinates. We propose to perform space quantization on $(x,y)$ to transform continuous position coordinates into neighborhood area classes. Each data sample now becomes $(\vec{s}, c, (x,y))$, where $c$ is a neighborhood area classes ID. Specifically, we divide $S$ into non-overlapping square grids with a side length of $\tau$. In practice, we set $\tau$ to be less than 0.2m. Then, we assign each grid neighborhoods a class ID $c$ and discard all classes without any data points. Thus, instead of using position coordinates as training labels, \noble uses neighborhood class as ground truth. During inference, \noble uses the predicted class to look up its neighborhood class's central coordinates and returns it as the prediction result.

\begin{figure}[htbp]
\centering
\includegraphics[trim= 0 0 0 0,clip,width=0.45\textwidth]{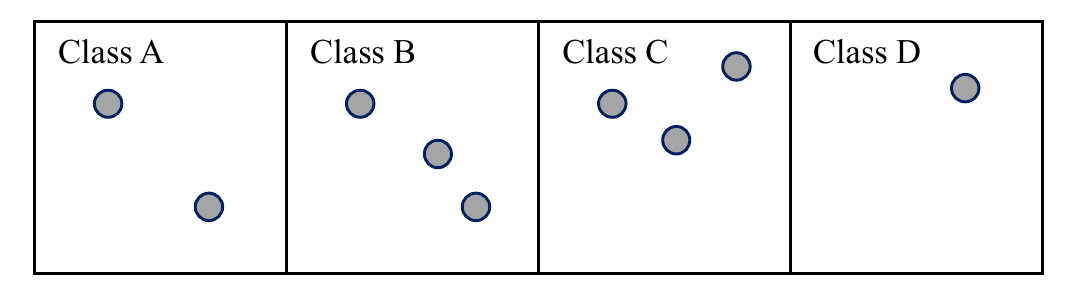} 
\caption{Consider this space for localization. Gray circles represent training data point locations. The whole space is quantized into four class and data points are labeled with its corresponding grid class ID.}
\label{fig:quantizeSpace}
\end{figure}


Our quantization approach exploits and approximates the ground truth closeness between data points in the output space without relying on Euclidean distance in the input space as neighborhood approximation. Moreover, assume a thorough training data sampling process over space $S$, our method eliminates inaccessible areas such as dead or irrelevant space from the output space because samples either cannot be or are not intentionally collected from those areas. For example, in Fig. \ref{fig:views-intro}, the middle area of top left buildings will not translate to any neighborhood classes as no data resides in that area.

Our space quantization enables us to solve the manifold regression problem with a fine-grained classification model. However, we have introduced one hurdle.  The classification problem is likely to suffer from class data sparsity. Since our grid is fine-grained, it is likely to contain very few training samples. We could increase $\tau$, or we could assign data samples with multiple classes, the ones that are adjacent to the real class. Moreover, we could also divide space $S$ into grid neighborhood of different length, $\tau$ and $l$ where $\tau < l$ . Each data sample now becomes $(\vec{s}, c, r (x,y))$ where $c$ denotes for the neighborhood classes ID of size $\tau$ and $r$ denotes neighborhood classes ID of size $l$. This formulation gives different levels of granularity of the output manifold. 

\vspace{-2mm}
\subsection{Why DNN Classification is Equivalent to Manifold Learning}

We will make the connection between manifold learning and our approach mathematically. To begin with, we introduce multidimensional scaling (MDS), a popular manifold learning algorithm, which has the objective:
$f(Z, \mathcal{X}) = \sum_{i=1}^n \sum_{j=1+i}^n ( ||z_i - z_j|| - ||x_i - x_j||)^2$ for $n$ points.
Essentially, MDS tries to learn embedding $Z$ on output manifold such that the pairwise relationships on input space are preserved. Close neighbors are encouraged to stay close in the reconstructed space and vise versa. 

In our formulation of \noble, we use binary cross-entropy loss function for multi-label classification, defined as
$J(h_c,\hat{h}_c) = \sum_{i=1}^n \sum_{c=1}^k -h_c \log (\hat{h}_c) - (1-h_c) \log (1-\hat{h}_c)$,
where $k$ is the number of classes, $n$ is the number of training data, $h_c \in \{0,1\}$ indicates the right class when $h_c = 1$, and $\hat{h}_c$ is the sigmoid function: $\hat{h}_c = (1 + \exp(-w_c^{\top} z_i))^{-1}$. Here, $w_c$ denotes the weight vector for class $c$ at the last layer, and $z_i$ denotes for the output from the second last layer for input $x_i$.  We focus our analysis on the last layer because the second last layer output can be interpreted as learned embedding for input features. From a manifold learning perspective, embedding from the last layer can be interpreted as reconstructed embedding. For simplicity, suppose $w$ and $z$ are normalized, we can rewrite $\hat{h}$ from an inner product to the Euclidean distance form as
$\hat{h}_c = (1 + \exp( \frac{1}{2}||w_c - z_i||^2 - 1))^{-1}.$

For a given $c$, minimizing the cross entropy loss will result in a setting such that $||w_c - z_i||$ for the true class is minimized (cf. false class is maximized). Consider $z_i$ as embedding given input $x_i$, $z_j$ as embedding given input $x_j$. If $x_i$, $x_j$ are near neighbors, then by our formulation, $x_i$, $x_j$ share same class label. Thus, the following holds for two embedding $z_i$ and $z_j$,
$||w_c - z_i||^2 \leq \lambda$ and $||w_c - z_j||^2 \leq \lambda$,
where $\lambda$ is a small constant. And by triangle inequality, we have
$||z_i - z_j||^2 \leq 2\lambda$. 
As we can see, $z_i$, $z_j$ is expected to be close, which resembles the objective function of MDS without considering the distance in the input space between $x_i$ and $x_j$.

We present \noble as a DNN based approach for localization that can utilize structure information. It should be noted that our evaluation measure is still position error (root mean square error) even though we transform the data into fine-grained classification inspired by manifold learning.  In the next two sections, we will use \noble on two orthogonal input signals, \wifi for the positioning task, and IMU for the tracking task.

%% file: Wifiexperiment.tex
\section{Application \wifi Localization}
In this section, we first present the detailed system design of \noble for \wifi fingerprinting localization. We conduct experiments on two representative indoor \wifi localization datasets: \UJI \cite{UJILOC}, the largest open-access dataset for indoor \wifi localization for large space multi-building setting, and IPIN2016~\cite{ipin2016} for small single building setting. 

\vspace{-2mm}
\subsection{System Architecture}

We follow the standard setup for \wifi fingerprint localization. Assume there are $W$ number of WAPs in the given space. During the offline phase, \wifi strength signal readings received from each WAP at each sampling location are recorded. Floor, building, longitude, and latitude are also recorded for each sample. Both \UJI and IPIN2016 are collected in such a way, and each sample can be represented as $(\vec{s}, b, f, (x,y))$. $\vec{s} = (s_1, s_2, \ldots, s_W)$, where $s_i$ denotes the RSS of $i$-th WAP, $b$ denotes building ID, $f$ denotes floor ID. Given the collected data, we perform output space quantization and convert each sample as $(\vec{s}, b, f, c, r,(x,y))$.  Apply the \noble multi-label classification formulation, our model takes $\vec{s}$ as inputs, and predict $(b, f, c, r)$. During inference, we use $c$ to look up the corresponding central coordinates, and output $(x_c, y_c)$ as position and calculate position error accordingly. One advantage of \noble is that we can naturally include floor/building classification tasks in our model without extra effort. Floor/building classification is a standard task for localization service. Current approaches utilize separate and independent models for position prediction and building/floor classification, creating extra overhead in real-world deployments. At the same time, from a manifold perspective, including floor/building as output is beneficial for the model to learn the reconstructed embedding because it gives useful information about geodesic neighborhood over the manifold structure.

\begin{figure}[htbp]
\centering
\includegraphics[width=0.4\textwidth]{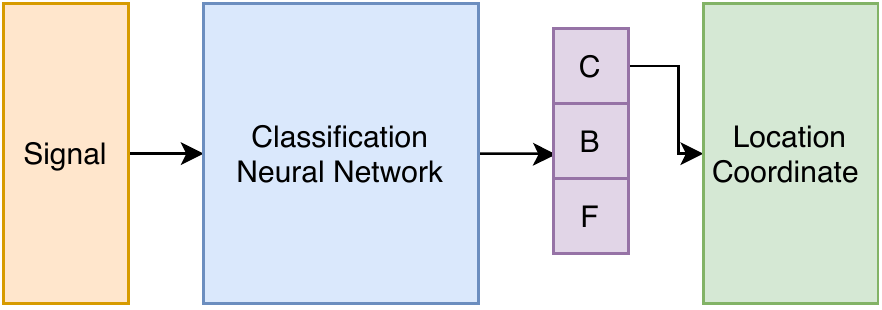}
\caption{Given wireless signal strength as input, \noble predicts multiple labels, which includes C for neighborhood class, B for building, F for floor. At inference time, \noble computes longitude and latitude coordinates based on the predicted neighborhood class.}
\label{fig:nobel_system_model}
\end{figure}
\vspace{-2mm}
We consider a two hidden layer feed-forward neural network that takes input vector $\vec{s} \in \mathbb{R}^W$. The hidden layer size is set to 128. We normalize the input vector and apply multi-hot encoding to the output class. We used hyperbolic tangent activation functions, Xavier initialization \cite{xavier}, and batch normalization \cite{batch-norm} for training our model. The overview of our system is shown in Fig. \ref{fig:nobel_system_model}.

\input{plots.tex}

\subsection{Performance Evaluation}
In our experiments, we first show that \noble achieves the best performance compared to all other approaches on the same datasets. Moreover, we set up three comparative models to demonstrate that \noble is aware of the output structure. We applied the best effort hyperparameter tuning for all methods.

\begin{table}[htbp]

\begin{center}
\footnotesize
\begin{sc}
\caption{\noble performance results on \UJI.}
\label{wifi_acc}
\begin{tabular}{lc}
\toprule
 & Classification Accuracy (\%)  \\
\midrule
Building  & 99.74         \\
Floor  & 94.25         \\
Quantize Class & 61.63 \\
\midrule
 & Position Error Distances (m) \\
\midrule
Mean   & 4.45  \\
Median  & 0.23  \\
\bottomrule
\end{tabular}
\end{sc}
\end{center}
\end{table}
\vspace{-2mm}

We calculate position error following the standard procedure: the Euclidean distance between predicted and true coordinates. For the \UJI dataset, the best mean error distance on the indoor localization ranking at IndoorLocPlatform website \cite{ipin2016} is 6.2 m, and the median is 4.63m. \cite{CNNLoc} reports a mean position error of 11.78 m, a building hit rate around 99\%, and a floor hit rate around 94\%. \cite{KS} reports a mean position error of 9.29m, a building hit rate around 99\%, and a floor hit rate around 91\%. As we can see in Table \ref{wifi_acc}, \noble achieves significantly smaller position error distances and at least comparable building and floor hit rate.

In order to evaluate the performance improvement from the perspective of structure awareness, we implement three comparison models: Deep Regression, Deep Regression Projection, and Manifold Embedding. Deep Regression takes the same input as \noble. It is the same network size as \noble. However, it is trained with mean square error as loss function and directly predicts coordinates in longitude and latitude. Deep Regression Projection is based on ~\cite{Chen_date}. Following Deep Regression, Deep Regression Projection projects the predicted coordinates to the nearest position on the map when the predictions do not lie on the map. Manifold Embedding utilizes Isomap and LLE to compute embedding from input signals. We built DNNs with two hidden layers that take the manifold embedding as input and output longitude and latitude coordinates. Manifold Embedding achieves the best performance when we set the embedding dimension at 400 for both Isomap and LLE. The performance results for the models mentioned above are shown in Table \ref{table_wifi_dis}.

\begin{table}[htbp]
\footnotesize
\begin{center}

\begin{sc}
\caption{Comparative distance (m) errors on \UJI.}
\label{table_wifi_dis}
\begin{tabular}{lcc}
\toprule
Model       & Mean & Median \\
\toprule
Deep Regression  & 10.17    & 7.84        \\
Regression Projection & 9.76  & 7.16 \\
Isomap Deep Regression & 11.01  & 7.56 \\
LLE Deep Regression & 10.05  & 7.43 \\
\bottomrule
\end{tabular}
\end{sc}
\end{center}
\end{table}
\vspace{-2mm}
Fig. \ref{fig:wifi-reg-pred}, \ref{fig:wifi-reg-proj-pred}, \ref{fig:wifi-reg-proj-pred}, and  \ref{fig:wifi-cls-pred}, are plots of predicted coordinates on the \UJI dataset.  \noble outputs the most structured prediction compared to the true floor plan. We see that deep regression outputs are spread out. From the satellite view in Fig. \ref{fig:views-intro}, we know that middle area of the top left building is not part of buildings; however, a considerable number of the deep regression outputs lie in this area. Manifold Embedding predicts fewer points in this area and is visually more structured compared to Deep Regression. This is as expected because Isomap Embedding is reconstructed with the aim to approximate output structure.  Also, Deep Regression Projection resembles the building structure because it eliminates prediction based on human-crafted maps.


\input{IMUFigure.tex}

On IPIN2016, \noble achieves an average error distance of 1.13m and a median average error distance of 0.046m, while the Deep Regression gives an average error distance of 3.83m. The best mean error distance on the indoor localization ranking at IndoorLocPlatform website \cite{ipin2016} is 3.71m. 
\vspace{-2mm}
\subsection{Energy Measurement:} We measure energy consumption on the Nvidia Jetson TX2 module. Using \UJI, the average running energy for each inference is 0.00518J, and the average latency is 2 milliseconds.

%% file: plots.tex
\begin{figure*}[htbp]
    \centering
	\begin{subfigure}[t]{0.23\textwidth}
	    \includegraphics[trim=60 60 60 60,clip,scale=.15]{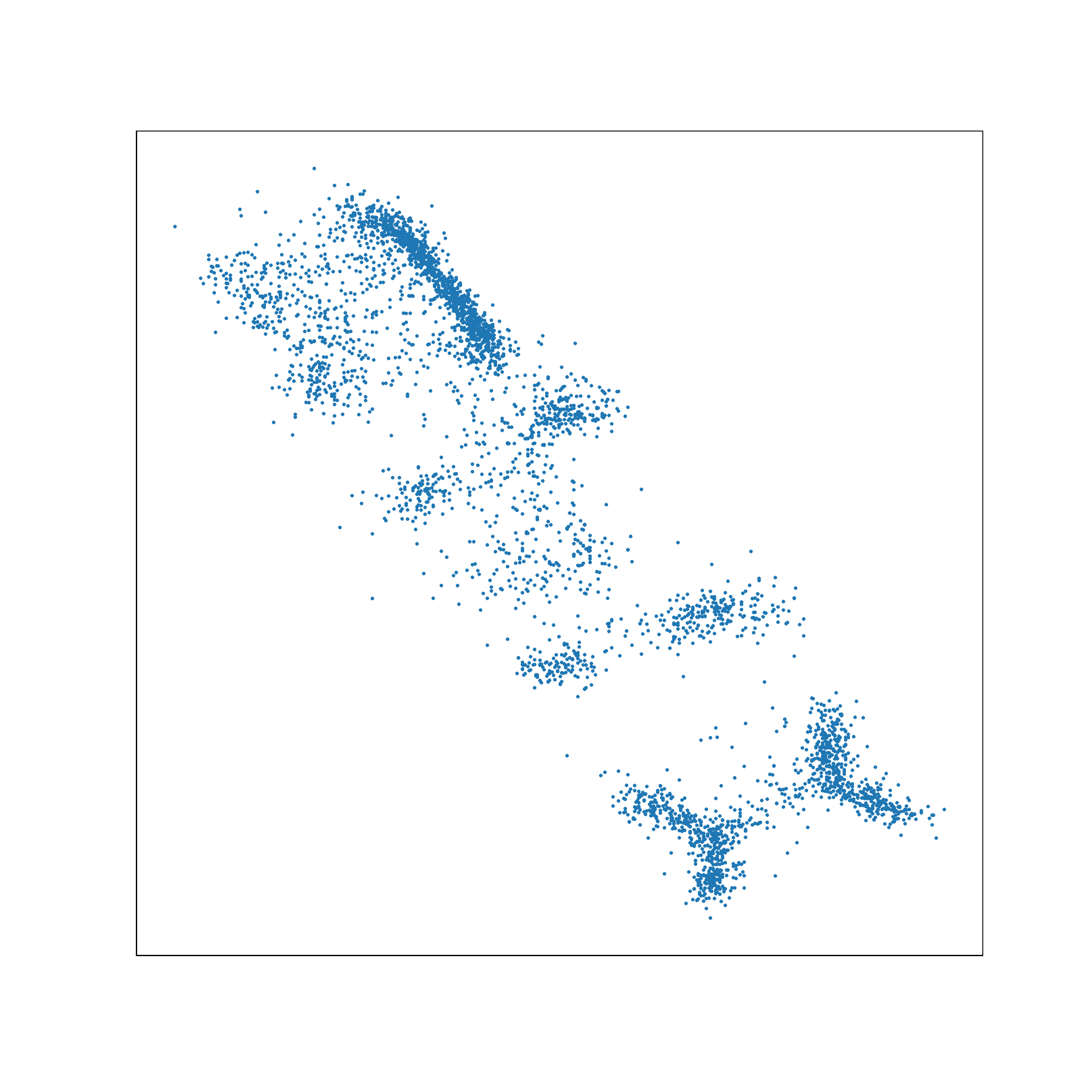}
	    \caption{Deep Regression}
	    \label{fig:wifi-reg-pred}
    \end{subfigure}
    \begin{subfigure}[t]{0.23\textwidth}
	    \includegraphics[trim=60 60 60 60,clip,scale=.15]{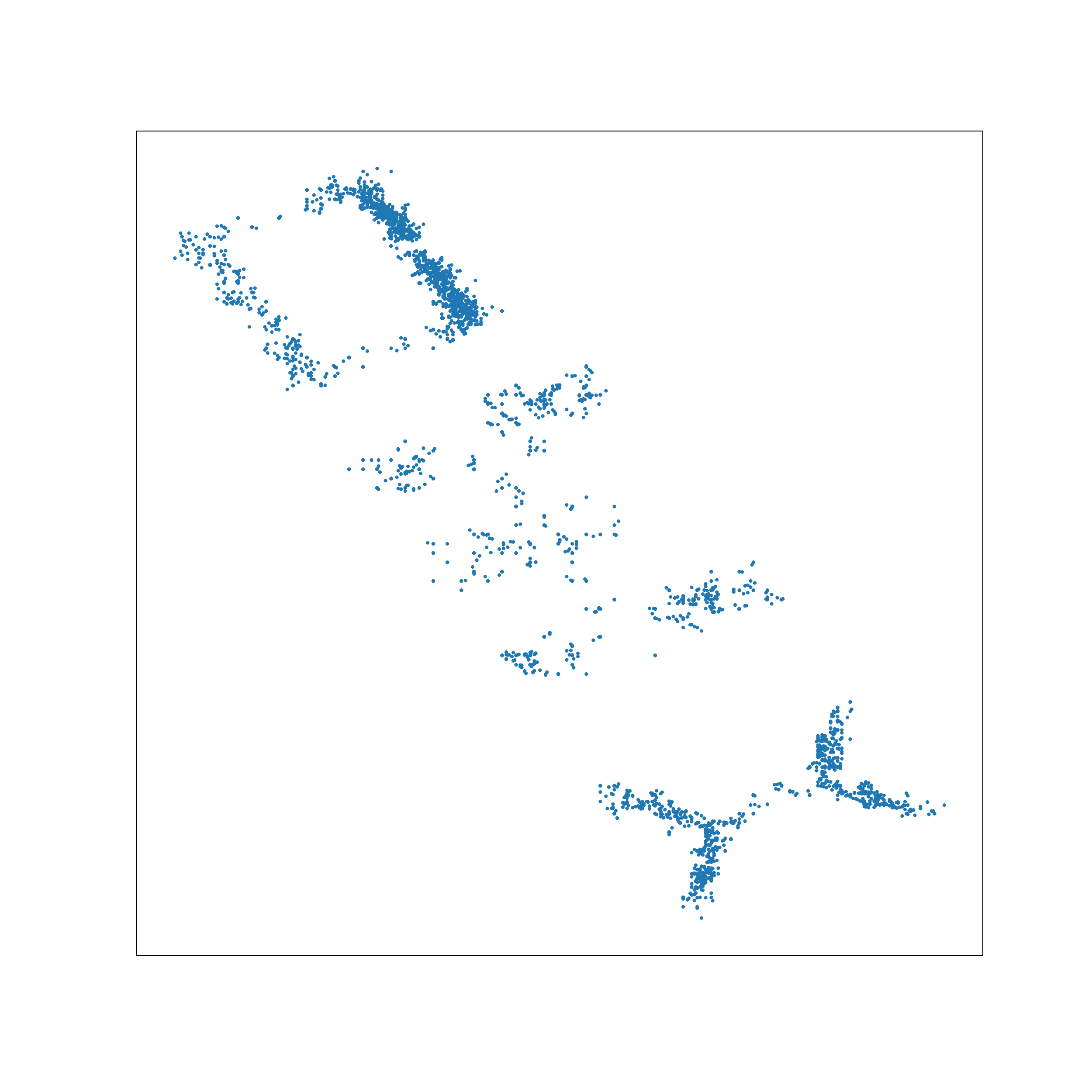}
	    \caption{Deep Regression Projection}
	    \label{fig:wifi-reg-proj-pred}
    \end{subfigure}
    \begin{subfigure}[t]{0.23\textwidth}
	    \includegraphics[trim=60 60 60 60,clip,scale=.15]{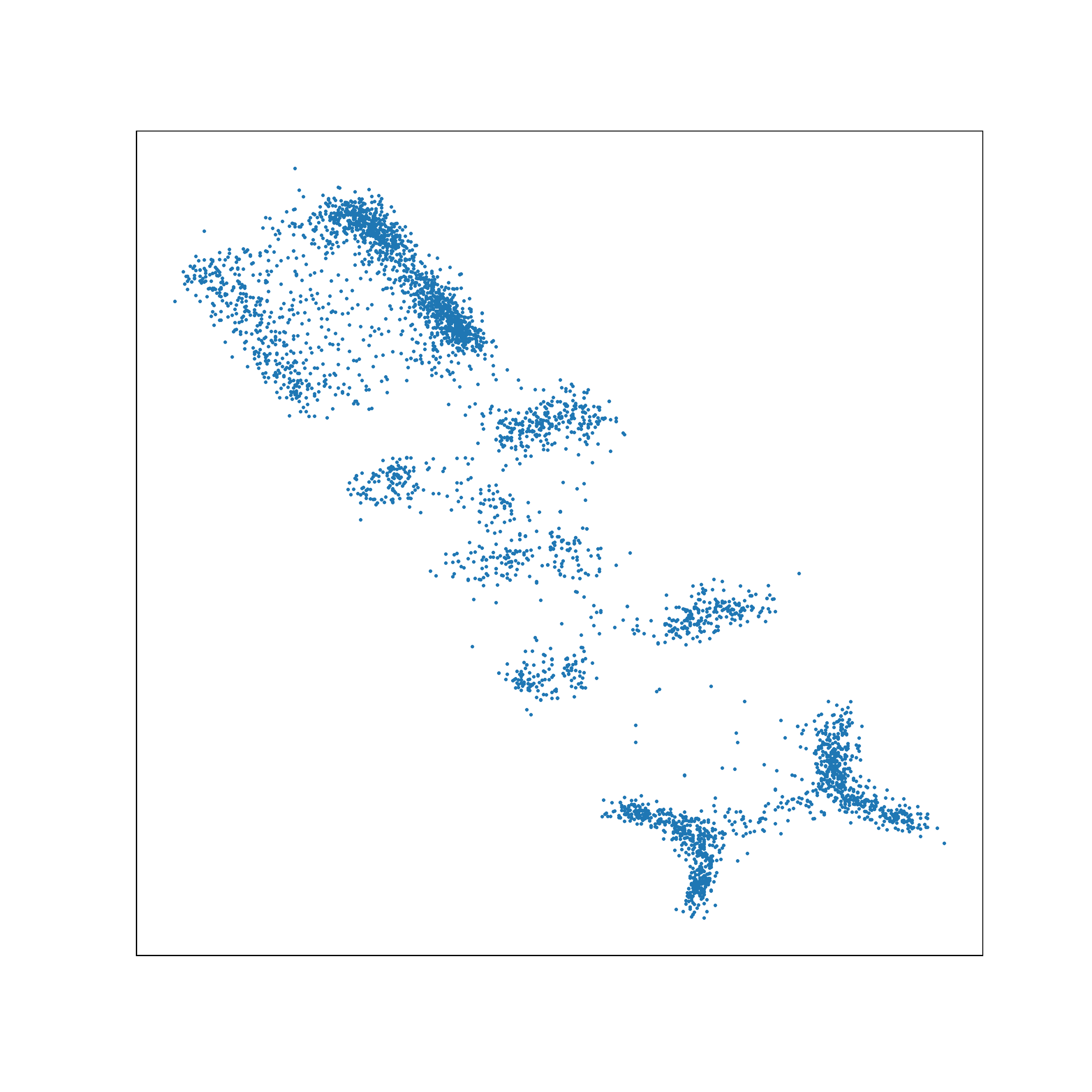}
	    \caption{Isomap Regression ($d$=400)}
	    \label{fig:wifi-iso-pred-400}
    \end{subfigure}
    \begin{subfigure}[t]{0.23\textwidth}
	    \includegraphics[trim=60 60 60 60,clip,scale=.15]{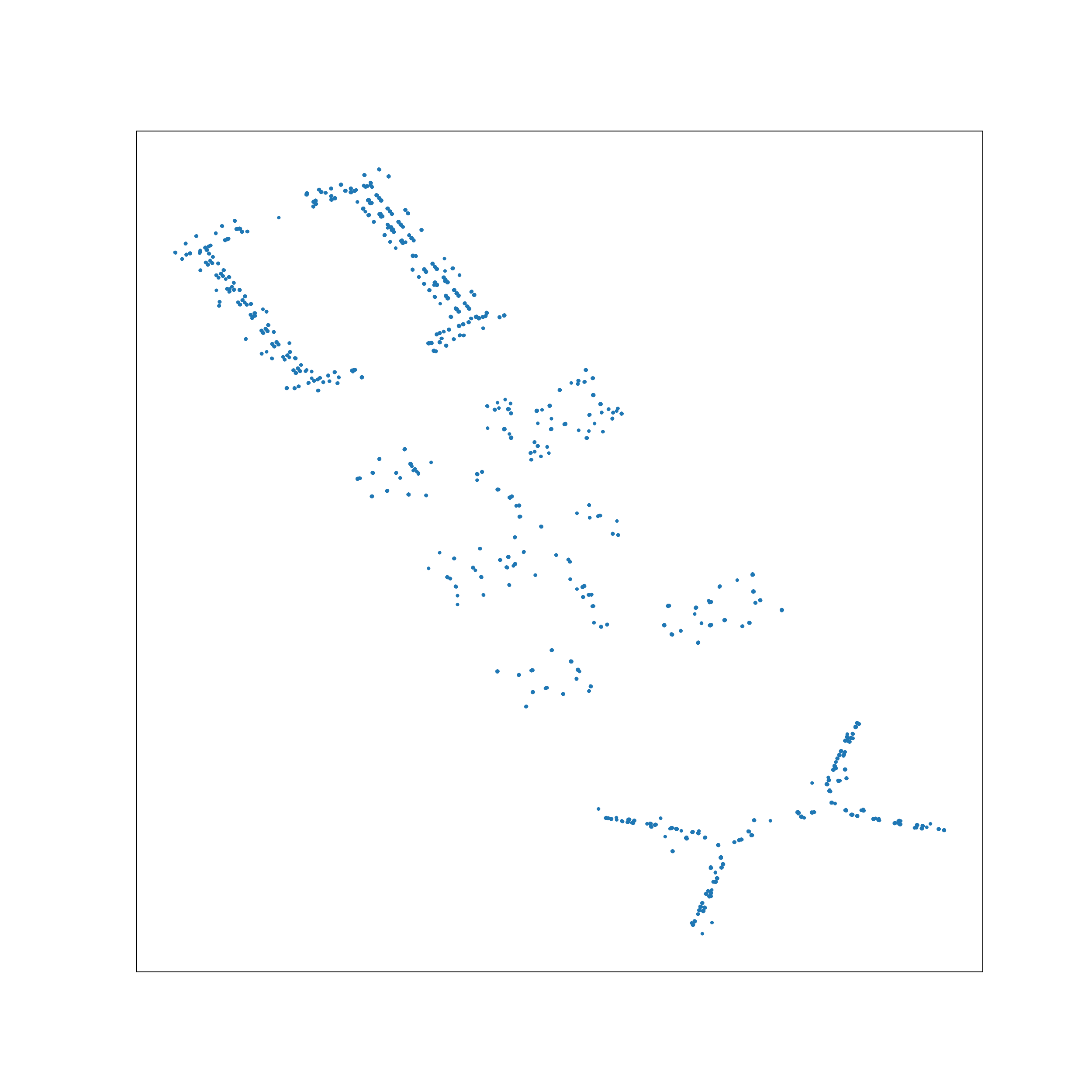}
	    \caption{\noble}
	    \label{fig:wifi-cls-pred}
    \end{subfigure}
    \caption{Plots of predicted coordinates from four models (labeled below each plot).}
\vskip -0.1in
\vspace{-2mm}
\end{figure*}

%% file: IMUFigure.tex
\begin{figure*}
    \centering
    \begin{subfigure}[t]{0.5\textwidth}
	    \includegraphics[trim=40 0 0 0, scale=.18]{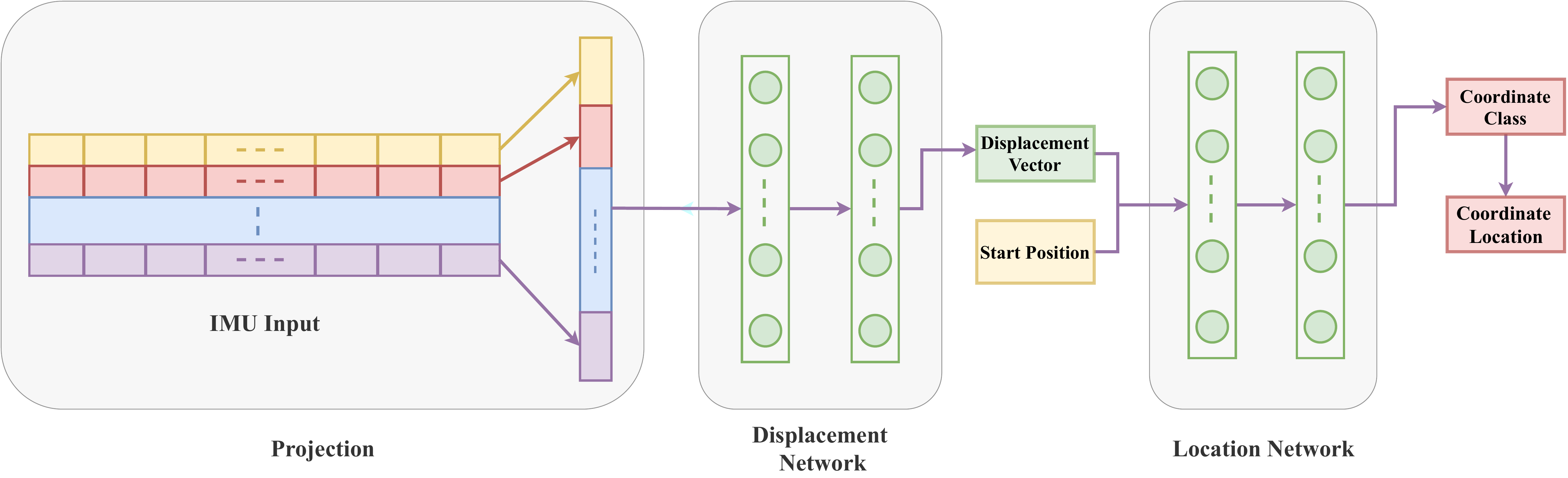}
	    \caption{}
	    \label{fig:IMU_model}
    \end{subfigure}    
	\begin{subfigure}[t]{0.15\textwidth}
        \includegraphics[trim=200 130 0 0, scale=.063]{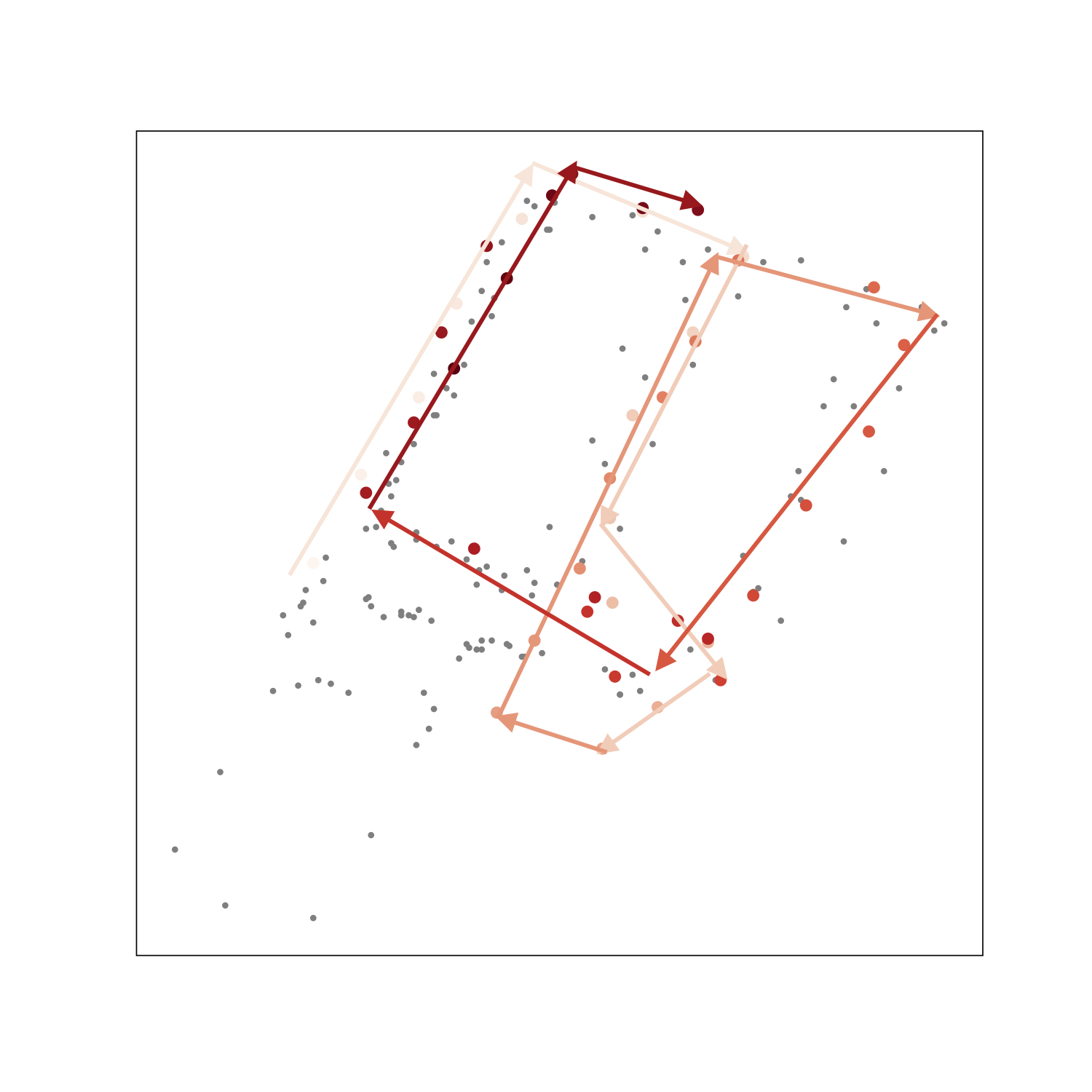}
        \caption{}
        \label{fig:imu-path}
    \end{subfigure}
	\begin{subfigure}[t]{0.15\textwidth}
	    \includegraphics[trim=80 60 60 60, clip, scale=.13]{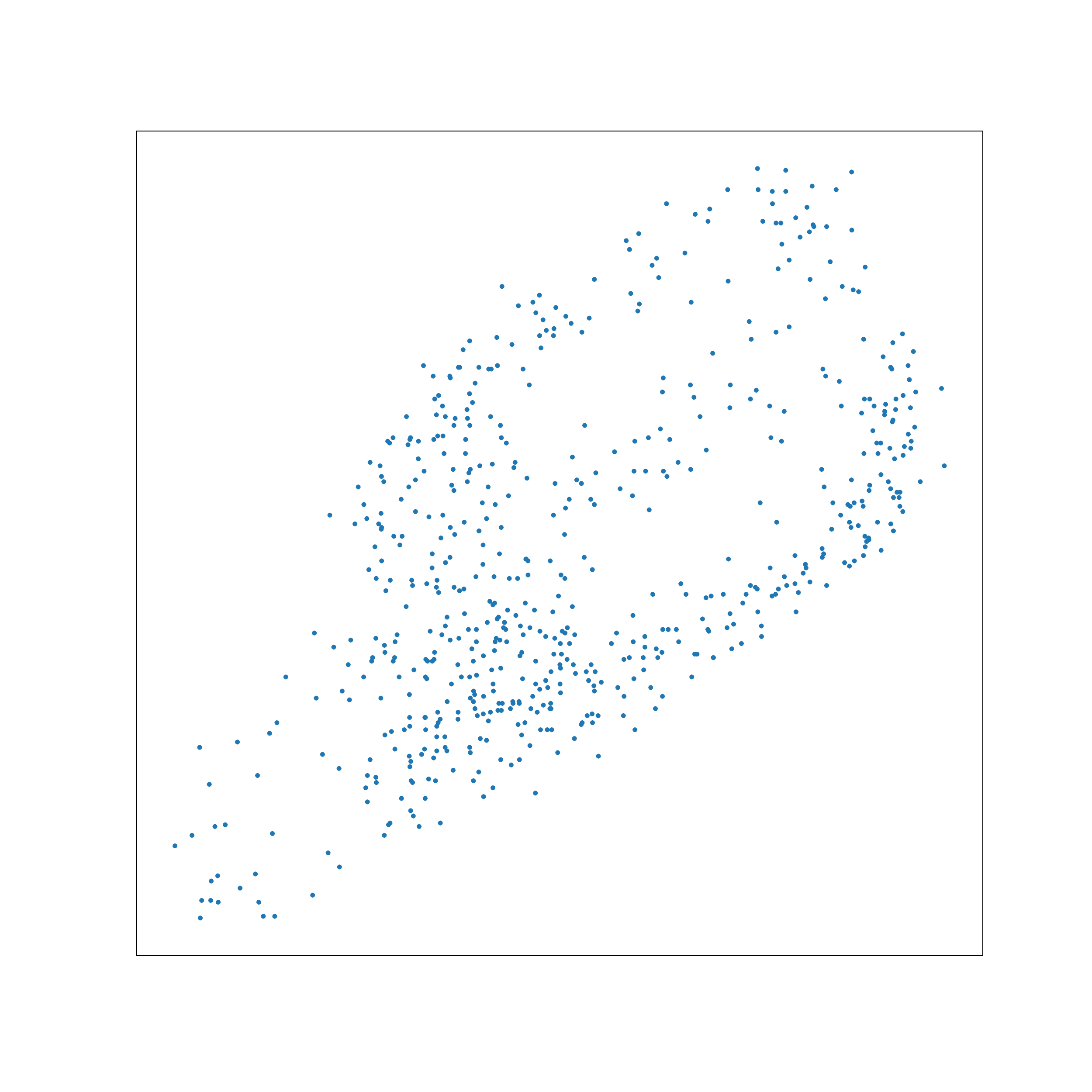}
	    \caption{}
	    \label{fig:IMU_Regr_pred}
    \end{subfigure}
    \begin{subfigure}[t]{0.15\textwidth}
	    \includegraphics[trim=60 60 60 60, clip, scale=.13]{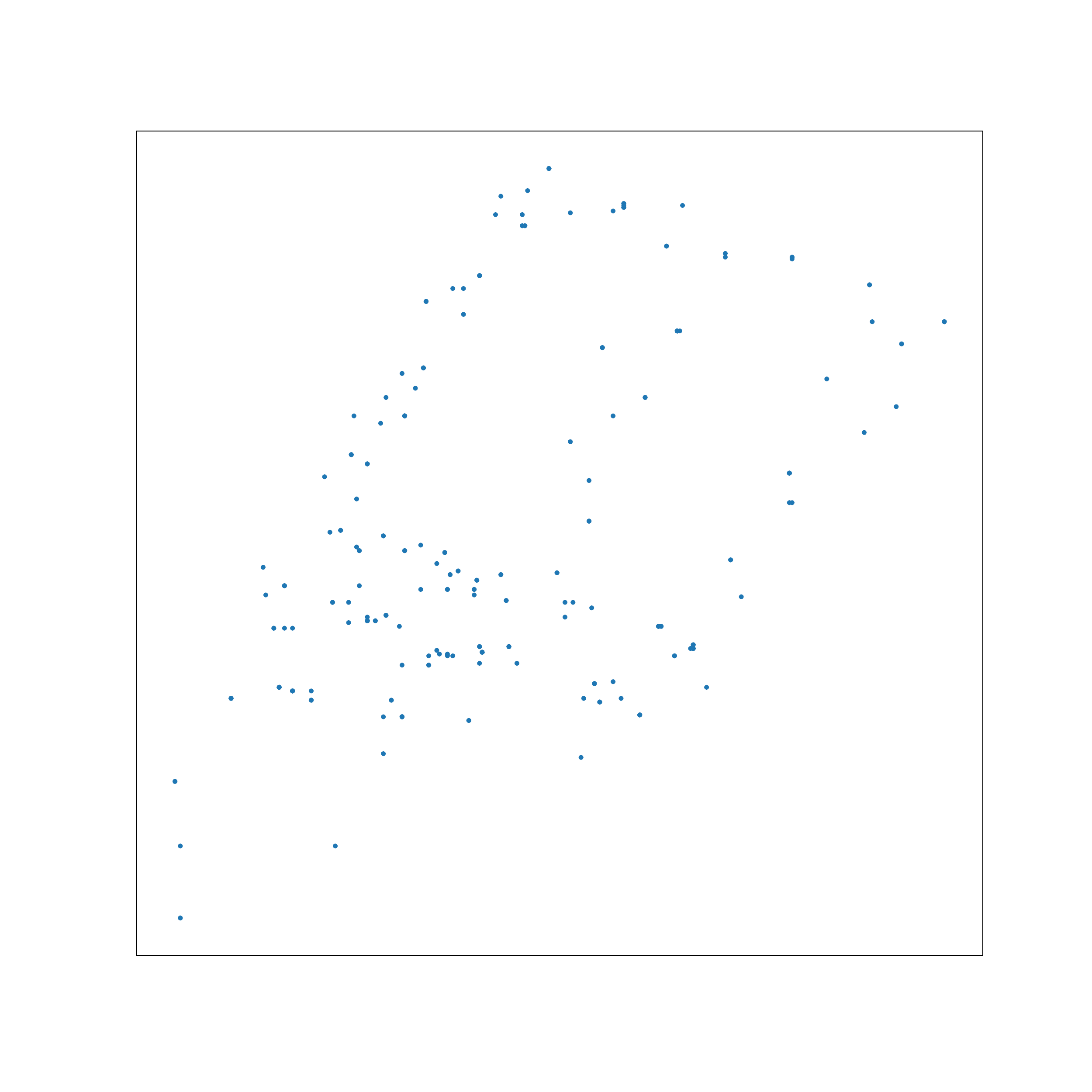}
	    \caption{}
	    \label{fig:IMU_Clas_pred}
    \end{subfigure}    
    \caption{(a) Network architecture for IMU localization. (b) User travel paths for testing. Color dots represent the sampling position along this path. Gray dots represents other sampling position in the dataset but not on this path. Changing of color represents the the travel sequence. (c) and (d) are IMU predicted coordinates for Deep Regression and \noble respectively.}
\end{figure*}

%% file: IMUExperiment.tex

\section{Application : Device Tracking Using IMUs }
In this section, we will discuss the detailed system design for device tracking using IMU signals. A user travels along a certain path, and a sequence of IMU data corresponding to this travel path is recorded. Given this sequence, we want to predict the user location at the end of this path. Without available public datasets, we collect labeled data over an outdoor space of 160m by 60m ourselves and show that \noble achieves accurate device tracking in terms of path ending position errors.
\vspace{-2mm}
\subsection{Data Collection}
We follow the standard setting of device tracking using IMU. We collect our data from two independent walks around an area of 160m by 60m on our university campus. The sampling frequency is around 50Hz, and the total walking time is around 1 hour and 15 minutes. There are in total 177 reference locations with GPS coordinates (longitude and latitude). Between each reference point, there are 768 readings for each inertial sensor on a single axis. We record 3-axis gyroscope, 3-axis accelerometer, and timestamps. We construct walking path as follows: (1) randomly choose a reference location as start position, (2) randomly choose a path length less than 50 and determine the end position accordingly, (3) concatenate IMU readings between starting and ending positions as the input. In total, we obtained 6857 paths, and we use 4389 for training, 1096 for validation, and 1372 for testing.
\vspace{-2mm}
\subsection{System Architecture}
 The input consists of two parts: (1) initial location coordinates $h_{start}$ and (2) a sequence of IMU signals $G = g_{1}, g_{2}, \ldots$, where $g_{i} \in \mathbb{R}^{d \times n}$. $d$ is the dimension of each inertial sensor readings and $n$ is the number of sensors. We perform output space quantization at $\tau = 0.4$m and assign neighborhood classes $c$ for path ending location. Following the \noble formulation, our model takes $(G, h_{start})$ as inputs, and predicts $\hat{c}$. Then, we calculate ending position in longitude and latitude based on predicted neighborhood class $\hat{c}$.
 
 Our system includes three main parts: (1) projection module, (2) displacement module, and (3) location module. The projection module takes $g_i$ and outputs an embedding in a lower dimension. Then, all projection embeddings are concatenated together. Each $g_i$ is multiplied by the same trainable projection weight. The concatenated embedding is passed into the displacement module, a two-layer feed-forward neural network that predicts the displacement vector of a user's travel path. This module is not environment-specific, and a trained module can be plugged into other models designed for location tracking in other environments. Taken projected embedding, the displacement network outputs a displacement vector $V \in \mathbb{R}^2$ for tracking on the 2-D plane or $V \in \mathbb{R}^3$ for 3-D tracking involving floors. The location network takes the resulting displacement vector and one-hot encoded starting location class, and outputs location class at the end of travel path. We used Xavier initialization \cite{xavier} and batch normalization \cite{batch-norm} for training. The overview of our system is shown in Fig. \ref{fig:IMU_model}.
\vspace{-2mm}
\subsection{Performance Evaluation}
\noble achieved a mean error distance of 2.52m and a median distance of 0.4m. \cite{Chen_date} iterative corrects prediction location at all turnings on the path and achieves an average error distance of 4.3m. LocMe \cite{7917859} reports a median of 1.1m position error on test-bed size of 70m by 100m by constantly correcting at elevators and walls. We could not test their method on our dataset as they did not open source their code. It is evident that incorporating of map knowledge is essential in these two previous works. However, both of these systems require human effort to transfer map knowledge into heuristic rules. 


Similar to our experiment on \wifi Localization, we implemented Deep Regression in order to demonstrate \noble's structure awareness. The results are shown in Table \ref{sample-table}. 
\begin{table}[htbp]
\vspace{-2mm}
\vskip -0.1in
\footnotesize
\caption{Position error distance (m) for IMU tracking.}
\label{sample-table}
\begin{center}
\begin{sc}
\begin{tabular}{lcc}
\toprule
      & Mean  & Median  \\
\midrule
Deep Regression Model  & 10.41     & 10.05      \\
\cite{Chen_date} & 4.3 & n/a\\
\noble & 2.52  & 0.4 \\
\bottomrule
\end{tabular}
\end{sc}
\end{center}
\vspace{-4mm}
\vskip -0.1in
\end{table}

Fig. \ref{fig:imu-path} shows information on the IMU dataset. From Fig. \ref{fig:IMU_Regr_pred}, it is evident that Deep Regression performs poorly on estimating the structural knowledge of the space since the predicted locations, blue dots, are scattered in the space. In contrast, \noble performs better in capturing the structural information since the predicted location points more closely resembles the space structure as seen in Fig. \ref{fig:IMU_Clas_pred} (cf. Fig. \ref{fig:imu-path}).

\vspace{-2mm}

\subsection{Energy Measurement}
We measured energy consumption on an edge computing device emulator, Nvidia Jetson TX2 module. For a testing path for around 8 seconds, \noble consumed around 0.08599J for inference calculation with a 5 milliseconds latency. Inertial sensors' energy cost is 0.1356J for 8 seconds, and the total energy consumption is approximately 0.22159J, which is 27$\times$ less than the GPS energy requirement 5.925J based on \cite{Chen_date}.

%% file: conclusion.tex
\section{Conclusion}
We propose a novel method for accurate localization and device tracking problem, {\bf N}{\em eighbor} {\bf Ob}{\em livious} {\bf Le}{\em arning} ({\bf \noble}), with the focus on the structure of the output space. We demonstrated that our formulation is essentially equivalent to manifold learning without approximation of local Euclidean distances in the input space. We applied \noble on two orthogonal applications, \wifi localization and IMU tracking, and showed a significant increase in localization accuracy.
